\documentclass[a4paper,12pt]{article}
\usepackage{epsfig}
\newcommand{\A}{\alpha}
\newcommand{\B}{\beta}
\newcommand{\G}{\gamma}
\newcommand{\D}{\delta}
\newcommand{\w}{\omega}
\newcommand{\F}{\frac}
\newcommand{\I}{\left}
\newcommand{\J}{\right}
\newcommand{\p}{{\partial}}
\newcommand{\n}{{\nabla}}
\renewcommand{\k}{{\hat{\f k}}}
\newcommand{\pau}{\F{\h{\f{\sigma}}}2}
\newcommand{\s}{\sigma}
\newcommand{\z}{\hat{\f z}}
\newcommand{\X}{\times}
\newcommand{\h}[1]{\hat{#1}}
\newcommand{\hS}{\hat{\f S}}
\newcommand{\sk}{\f m_{\hat k}}
\newcommand{\e}{\epsilon}
\newcommand{\be}{\begin{equation}}
\newcommand{\ee}{\end{equation}}
\newcommand{\bea}{\begin{eqnarray}} 
\newcommand{\eea}{\end{eqnarray}} 

\newcommand{\f}[1]{\mbox{\boldmath$#1$}}
\newcommand{\Ham}{{\cal{ H}}}
\newcommand{\R}{{\mathcal{ R}}}
\newcommand{\ba}{\begin{array}} 
\newcommand{\ea}{\end{array}} 
 
\newcommand{\const}{\mathop{\rm const}\nolimits} 
\newcommand{\Sp}{\mathop{\rm Sp}\nolimits}

\renewcommand{\Im}{\mathop{\rm Im}\nolimits} 

\title{Spin Dynamics of a Fermi Liquid in an Electric Field}
\author{P. L. Krotkov\footnote{E-mail: krotkov@itp.ac.ru.}\\\em L.D. Landau Institute for Theoretical Physics, Russian
Academy of Sciences, \\\em117334 Moscow, Russia}
\begin{document}
\large
\maketitle
\begin{abstract}
The influence of an external electric field on the spin dynamics of an electrically neutral Fermi
liquid is considered, the mechanism of such an influence being the relativistic spin-orbital
interaction. As a result,  Leggett's equations for the spin dynamics of weakly polarized Fermi liquids are
generalized to the  case of non-zero external electric field.
In addition, we obtained the transverse spin dynamics equation for strongly spin-polarized liquids
in an electric field at zero temperature. In both situations covariant derivatives depending on the
electric field are shown to be substituted for spatial gradients
in line with the SU(2) gauge invariance of the microscopic Hamiltonian.
The new equations are applied to the study of spin flow along a channel,
where an electric field is found to bring about
an additional phase shift of the order of magnitude of the phase shift
in superfluid
$^3$He-$B$ but growing with time.
\end{abstract}
\section{Introduction}
The advances in experimental technique in
recent years made feasible the observation of effects due to electric field on the  motion of
electrically neutral superfluid $^3$He~\cite{Bor}. Experiments on superfluid
$^3$He-$B$ in an electric field, using the spin current Josephson effect, are under way at the moment~\cite{Dm}. The
influence of an external electric field on superfluid $^3$He has also been treated
theoretically in scientific literature~\cite{M&M,M,M&V}.
However, a similar investigation for a normal Fermi liquid is lacking. It is this question which is attempted to be
answered by the present paper.

There have been predicted two competitive mechanisms of the influence of an electric field on the spin dynamics of the
superfluid $^3$He~\cite{M&V}, viz., (i)
slight deformation of the electronic shells of $^3$He atoms by the gradient of the order parameter
and (ii) spin-orbital coupling $(\f\mu\X\f p)\f E/mc$  of the electric field with the moving magnetic
moments of the $^3$He nuclei, where $\f\mu$ is the magnetic moment of $^3$He atom.
In the present paper this latter relativistic effect is incorporated into spin dynamics equations of
a \emph{ normal} Fermi liquid. The result complies with the gauge invariance arguments developed in
Ref.~\cite{M&V}.

In superfluid $^3$He-$B$ in an experiment with fixed spin supercurrent along a channel of length $L$, connecting two
reservoirs with homogeneously precessing magnetizations, the supercurrent, induced by an electric field, leads to an
additional phase shift of the order of
\be
\D\A\sim\F{2\mu E}{\hbar c}L\sim 10^{-4}\mbox{  rad}
\ee
for $E\approx 3\cdot 10^{4}$ V/cm and $L\approx1$ cm. According to~\cite{Bor}, this is an experimentally detectable quantity.

In a normal Fermi liquid a phase shift of the same order is demonstrated
to arise and grow with time in a similar one-dimensional spin flow
geometry.

The set-up of the paper is as follows. The second section is dedicated to introducing the SU(2) gauge invariant theory
of the interaction of
electromagnetic field with a Fermi liquid. In the third section the weakly polarized Fermi liquid is
considered on the basis of the
modifications to the Landau theory resulting from the appearance of an electric field. Although
this section does not overtly leans on the gauge invariance, the outcome agrees with what one
expects from Sec. II. The next section is given over to the study of one-dimensional spin flow as an example of
application of the new equations.

The invariance properties of the microscopic Hamiltonian are employed in treating
strongly spin-polarized liquids at zero temperature in Sec.~V. The last section sums up the main results.
\section{Gauge Invariance}
In the SU(2)-gauge-invariant wording of Ref.~\cite{M&V} electric and magnetic fields are introduced
uniformly as components of a gauge field $\f A_\mu$, where  $\mu=0$, $1$, $2$, $3$:
\be
g\f A_0=g\f H\equiv\f\w_{\mathrm{L}},\;\;\;\;gA^\A_i=-\F{g}{c}e_{\A ik}E_{k}.
\ee
Here $g$ is the gyromagnetic ratio for $^3$He nuclei, $\f\w_{\mathrm{L}}$ is the Larmor frequency.

The Hamiltonian of particles with the mass $m$ and the magnetic moment $\mu=\hbar g/2$ in an external
electromagnetic field in the second quantization representation is
\be
\Ham=\int\psi^+(\f r)\I[\F{\h p_i^2}{2m}-\h{\f\mu}\I(\f H-\F1{mc}\h{\f p}\X\f E\J)\J]\psi(\f r) \;d^3 \f r,
\ee
where $\h{\f\mu}=\mu\h{\f{\s}}$  is the magnetic moment operator. The Hamiltonian can be rewritten equivalently
using the gauge field $\f A_\mu$:
\be
\Ham=\int\psi^+(\f r)\I[\F{\I(\h p_i-\h{\f\mu}\f A_i\J)^2}{2m}-\h{\f\mu}\f A_0 \J]\psi(\f r) \;d^3
\f r.\label{Ham}
\ee

The last expression is invariant against rotations of the spin space if only they are accompanied by the
corresponding change of the gauge field:
\be
\psi\to\R\psi,\;\;\;\;g\f A_\mu\to g\f\R \f A_\mu +\f\w_\mu,
\ee
where $\R=\exp(-i\f\theta\h{\f\s}/2)$ is a 2D finite rotation matrix on the angle $\theta$ around the axis
$\h{\f\theta}$ and $\f\R(\f\theta)$ is a 3D rotation matrix  corresponding to $\R$. The frequency tensor of this
rotation $\f\w_\mu$ is defined as  $-i\f\w_\mu\h{\f\s}/2=\R^+\p_\mu\R$, and the gradients are
designated as $\p_\mu$, where $\p_0=\p_t$, $\p_i=\nabla_i$.

This symmetry allows one to refer to $\f A_\mu$ as the gauge field. Quite generally, gauge fields appear when the invariance
of a Hamiltonian against some \emph{ local} gauge symmetry is stipulated.  Nevertheless, it might be worth mentioning that
this symmetry is somewhat formal since it is valid  only for the Hamiltonian of $^3$He atoms and not for the
electromagnetic field.

The SU(2) gauge invariance of the Hamiltonian calls for all observables to be gauge invariant as well.
Hence all space-time derivatives can enter spin dynamics equations only in combinations with the gauge field, i.e. in
the form of covariant derivatives:
\be
D_\mu\f X=\p_\mu\f X+g\f A_\mu\X\f X.
\ee

As we will see later,  this conjecture is correct.
\section{Weak Polarizations}
Spin dynamics equations in the case of weak polarization  have been formulated by Silin~\cite{Silin} and
Leggett~\cite{Leg} in the framework of the Landau theory~\cite{Landau}. Leggett showed that the
first two harmonics of the distribution
function, viz., the spin and the spin current densities, decouple from the rest for slow enough spatial
variations. The  purpose of this section is to generalize the Leggett
equations  to the case of an electric field present.

The Landau theory of Fermi liquids operates with phenomenologically defined quantities. So it is not convenient to
literally use the gauge invariance approach discussed above. We will set this approach aside until Sec.~V and
formulate the theory in the usual terms of the quasiparticle distribution matrix $\h n_k$
and the quasiparticle energy  matrix $\h\e_k$.

In the absence of fields the equilibrium values of $\h n_k$ and $\h\e_k$ are
\be
\h\e^0_k=\xi_k\h 1,\;\;\;\;\h n^0_k=f_0(\h\e^0_k).\label{equil}
\ee
Here $f_0(\xi)=[\exp(\xi/T)+1]^{-1}$ is the Fermi distribution function and  $\xi_k=\hbar^2
k^2/2m^*-\e_{\mathrm{F}}$, where $m^*$ is the quasiparticle mass.

According to Landau's ideas the departure from equilibrium distribution
$\D\h n_k$ changes the quasiparticle energy along with external fields:
\be
\D\h\e_k=\D\h\e_k^{\mathrm{ext}}+\sum_{k'}\Sp'[f_{kk'}\D\h n_{k'}], \label{energy}
\ee
where $\D\h\e_k^{\mathrm{ext}}$ is the energy of a  quasiparticle in
 external fields, $f_{kk'}=N_{\mathrm{F}}^{-1}(F^{\mathrm{s}}_{kk'}\h 1\h 1'+F^{\mathrm{a}}_{kk'}\h{\f\s}\h{\f\s}')$ is the
quasiparticle interaction and $N_{\mathrm{F}}=m^*k_{\mathrm{F}}/\pi^2\hbar^2$ is the density of states on the Fermi surface.

In our case $\D\h\e_k^{\mathrm{ext}}$ is the Zeeman energy acquired by a quasiparticle in
a magnetic field:
 \be
 \D\h\e_k^{\mathrm{ext}}=-\h{\f\mu}\f H= -\F{\hbar\f\w_{\mathrm{L}}}2\h{\f\s}.\label{ext0}
 \ee

As the microscopic Hamiltonian~(\ref{Ham}) indicates, an electric field alters the equilibrium (kinetic) part
$\h\e^0_k$  of the energy matrix $\h\e_k$ by shifting  all the momenta of quasiparticles:
\be
\h\e^0_k\to\h\e^0_k+\D\h\e^0_k, \qquad \mbox{where } \D\h\e^0_k=-v_{{\mathrm{F}}i}\h{\f\mu}\f A_i.\label{amend1}
\ee

As a consequence, the equilibrium distribution $\h n^0_{k}\equiv f_0(\h\e^0_{k})$ will also change: $\h n^0_{k}\to \h
n^0_{k}+({df_0}/{d\xi})\D\h\e^0_{k}$, which, in its turn, will alter the departure from equilibrium $\D\h n_{k}$ defined as $\D\h n_{k}=\h
n_{k}-\h n^0_{k}$:
\be
\D\h n_{k}\to\D\h n_{k}-\F{df_0}{d\xi}\D\h\e^0_{k}.\label{amend2}
\ee

The amendment in $\D\h n_{k}$ through relation~(\ref{energy}) will change $\D\h\e_{k}$:
\be
\D\h\e_{k}\to
\D\h\e_{k}+\sum_{k'}\Sp'[f_{kk'}\I(-\F{df'_0}{d\xi'}\J)\D\h\e^0_{k'}]=\D\h\e_{k}+\F{F_1^{\mathrm{a}}}3\D\h\e^0_{k}.
\label{amend3}
\ee

Now let  introduce macroscopic quantities. Spin  and spin current densities are defined according to the formulas:
\be \f S=\F{\hbar}2\sum_k\Sp  [\h{\f\s}\h n_k],\;\;\;\;\f J_i=\F12\sum_k\Sp[\h{\f\s}\h n_k\n_{k_i}\h\e_k].\ee

Both of these expressions in equilibrium in the absence of external field are equal to zero. Expanding them to within the first order
in departures from equilibrium yields
\bea
\f S&=&\hbar\sum_{k}\Sp  [\h{\f\s}\D\h n_k],\label{S}\\
\f J_i&=&\F12\sum_k\Sp[\h{\f\s}\I(\D\h n_k\n_{k_i}\h\e_k^0+\h
n_k^0\n_{k_i}\D\h\e_k\J)]=\F{\hbar}2 \sum_{k}v_{{\mathrm{F}}i}\Sp  [\h{\f\s}\I(\D\h n_k-\F{d f_0}{d\xi}\D\h\e_k\J)],\label{J}
\eea
where the Fermi velocity $v_{{\mathrm{F}}i}=\hbar k_i/m^*$ and we took the second term in~(\ref{J}) by parts. This term describes
\emph{ backflow} (cf.~\cite{N}). The expression in parentheses in~(\ref{J}) can be easily seen to represent the
departure from the \emph{ local} equilibrium
$\h n_k^{\mathrm{loc}}=f_0(\h\e_k^0+\D\h\e_k)\approx f_0(\h\e_k^0)+(df_0/d\xi)\D\h\e_k$.

When an electric field is switched on both spin  and spin current densities in equilibrium will still be zero. This
follows from the fact that the amendments to their magnitudes brought about by $\D\h n^0_k$ and $\D\h\e^0_k$ are described by eqs.~(\ref{S})
and (\ref{J}) respectively. But these expressions vanish when $\D\h n^0_k$ and $\D\h\e^0_k$ are substituted for $\D\h
n_k$ and $\D\h\e_k$. Thus an external electric field produces no spin current in equilibrium.
We will exploit this fact later when writing out the collision integral in the kinetic equation in the
$\tau$-approximation.

Substituting the expression~(\ref{energy}) along with the renewed $\D\h n_k$ (\ref{amend2}) and $\D\h\e_k$
(\ref{amend3}) into~(\ref{J}) yields
\be
\f J_i=\F{\hbar N_{\mathrm{F}}}4\langle v_{{\mathrm{F}}i}\Sp[\h{\f\s}\D\h{\e}^{\mathrm{ext.eff.}}_k]\rangle_{\h k}+\F{\hbar}2\I(1+\F{F_1^{\mathrm{a}}}{3}\J)\sum_k v_{{\mathrm{F}}i}\Sp  [\h{\f\s}\D\h
n_k],\label{JJ}
\ee
where the corrections due to electric field to $\D\h n_k$ and $\D\h\e_k$ can be conveniently incorporated into an
effective external-fields energy
\be
\D\h{\e}_k^{\mathrm{ext.eff.}}=-\F{\hbar\h{\f \s}}2\I(\f\w_{\mathrm{L}}+v_{{\mathrm{F}}i}\I(1+\F{F_1^{\mathrm{a}}}3\J)g\f A_i\J)\label{ext}
\ee

 On plugging this into~(\ref{JJ}), we obtain the following  expression  for the spin current:
\be
\f J_i=\f J_i^0+\hbar\I(1+\F{F_1^{\mathrm{a}}}3\J)\sum_k v_{{\mathrm{F}}i}\f\s_k,
\ee
where $\f\s_k={\scriptstyle \F12}\Sp  [\h{\f\s}\D\h n_k]$. As compared to the case of no electric field~\cite{Leg}, the
expression for the spin current contains an additional term
\be
\f J_i^0=\F{\mu\rho\hbar}{2m^*c}e_{\A ik}E_k\I(1+\F{F_1^{\mathrm{a}}}3\J)=-\F{\chi_{\mathrm{n}}}{g^2}\F{w^2}3 g\f A_i,
\ee
where  $\rho=k_{\mathrm{F}}^3/3\pi^2$ is the liquid's density, $w^2=v_{\mathrm{F}}^2(1+F^{\mathrm{a}}_0)(1+F^{\mathrm{a}}_1/3)$ and
$\chi_{\mathrm{n}}=g^2\hbar^2 N_{\mathrm{F}}/4(1+F^{\mathrm{a}}_0)$ is the normal Fermi-liquid susceptibility.
This term appeared previously in the spin supercurrent in superfluid Fermi liquid~\cite{M&M}.

We now proceed to the derivation of the dynamics equations themselves.

Let expand $\h n_k$
and $\h\e_k$  in the basis of  the identity and the Pauli matrices:
\be
\h n_k=f_k\h 1+\f\s_k\h{\f\s}, \qquad\h\e_k=\e_k\h
1+\f\e_k\h{\f\s}.
\ee

  In the weak-polarization case, where spin polarization is a very small fraction of the total number of spins in the
liquid, $f_k$ and $\e_k$ may be put equal to their equilibrium values~(\ref{equil}) (see~\cite{Leg}). Then from
eqs.~(\ref{energy}),  (\ref{ext0}), (\ref{amend1}) and (\ref{amend3}) we obtain
\be
\f\e_k=-\F{\hbar}2\I(\f\w_{\mathrm{L}}+v_{{\mathrm{F}}i}\I(1+\F{F_1^{\mathrm{a}}}3\J)g\f
A_i\J)+\F2{N_{\mathrm{F}}}\sum_{k'}F^{\mathrm{a}}_{kk'}\f\s_{k'}.
\ee

Note that here  the electrical-field-dependent corrections to
$\h\e_k^0$ and  $\D\h\e_k$ also appear in effect as if it is the external energy $\D\h\e^{\mathrm{ext}}_k$ which contained them.
What we mean is that the above
expression could be obtained from the expression (\ref{energy}) with the \emph{effective} external energy (\ref{ext})
\emph{without} allowing for the corrections  (\ref{amend1}) and (\ref{amend3}).
It is tempting therefore to attribute the term in (\ref{ext}) that depends on electric field
to an external spin-orbital energy. Nevertheless, since it contains Fermi-liquid constants this term is not, strictly speaking, an
energy in an external field. Quite the reverse, it appeared from corrections to the kinetic energy. Thus we conclude
that the parallel drawn  is  formal.

The Silin-Leggett~\cite{Leg} kinetic equation for the spin part $\f\s_k$ of $\h n_k$ is  (for brevity we omit
$k$-indices)
\be
\p_t\f\s-\F2\hbar\f\e\X\f\s+v_{{\mathrm{F}}i}\n_i\I(\f\s-\F{d f_0}{d\xi}\f
\e\J)=(\p_t\f\s)_{\mathrm{coll}}. \label{kin}
\ee
Note the wrong sign before the second term in~\cite{Silin,Leg}.

Defining the averaged distribution
\be
\sk(\f r,t)=\F{N_{\mathrm{F}}}2\!\int \!d\xi\;\f\s_k(\f r,t),
\ee
we can further reduce the kinetic equation to
\bea
D_t\sk&-&\F{4}{\hbar N_{\mathrm{F}}}\langle F^{\mathrm{a}}(\k\k')\f m_{\hat k'}\rangle_{\hat k'}\X\sk+v_{{\mathrm{F}}i}\n_i\I(\sk+\langle
F^{\mathrm{a}}(\k\k')\f m_{\hat k'}
\rangle_{\hat k'}\J)+\nonumber\\
&+&v_{{\mathrm{F}}i}\I(1+\F{F_1^{\mathrm{a}}}3\J)g\f A_i\X\sk-\F{\hbar N_{\mathrm{F}}}{4}v_{{\mathrm{F}}i}\n_i\I(\f\w_{\mathrm{L}}+v_{Fj}\I(1+\F{F^{\mathrm{a}}_1}3\J)g\f A_j\J)=(\p_t\sk)_{\mathrm{coll}}.
\eea

Following Leggett, we seek after a solution in the form of the sum of the
zeroth and first harmonics of $\sk$:
\be
\hbar\sk=\f S+\F3{v_{\mathrm{F}}}\I(1+\F{F_1^{\mathrm{a}}}3\J)^{-1}\hat k_i(\f J_i-\f J_i^0).
\ee
The higher harmonics can be proved to decouple from the first two if only the
characteristic spatial scale $\lambda$ is greater than the mean free path \emph{ or}
the molecular field length~\cite{Leg}:
\be
\lambda>\min\I\{v_{\mathrm{F}}\tau, \F{v_{\mathrm{F}}}{\kappa\w_{\mathrm{L}}}\J\}.
\ee
Here $\kappa=(F^{\mathrm{a}}_1/3-F^{\mathrm{a}}_0)/(1+F^{\mathrm{a}}_0)$. This constraint coincides with that for
the case of zero electric field because the ``electric length'' $l_E=(gE/c)^{-1}$ appearing as a new
scale in the problem is much greater than all the above mentioned scales (for an electric field as strong as $E=3\cdot 10^4 $V/cm
the ``electric length'' shortens only to $l_E=10^6 $cm).

When the mean free
path is shorter than the molecular field length (which obviously is equivalent to the inequality
$\kappa\w_{\mathrm{L}}\tau<1$) we deal with the hydrodynamic
regime. The opposite case corresponds to the collisionless regime.

After some algebra we obtain a set of two equations:
\bea
D_t\f S+D_i\f J_i&=&0, \label{s}\\
D_t\f J_i-\p_t\f J_i^0+\F{w^2}3 D_i(\f S-\f S^{\mathrm{eq}}) +
\kappa\F{g^2}{\chi_{\mathrm{n}}}\f S\X\f J_i&=&-\F{\f J_i}
{\tau_1}.\label{j}
\eea
Here $\f S^{\mathrm{eq}}=\chi_{\mathrm{n}}\f\w_{\mathrm{L}}/g^2$ is the equilibrium magnetization, $\tau_1=\tau/(1+F^{\mathrm{a}}_1/3)$. As
compared to the Leggett equations, all the spatial derivatives  in~(\ref{s}), (\ref{j}) are replaced with
the covariant ones. The term $\p_t\f J_i^0=(\chi_{\mathrm{n}}/g^2)(w^2/3)(\nabla_i\f\w_{\mathrm{L}}-\f\nabla \w_{\mathrm{L}i})$ is non-zero only for an electric field
varying in time. We will assume this is not the case in what follows.

First, from now on we make use of the units wherein $\chi_{\mathrm{n}}=g^2$. Next, we transform to the frame, rotating with the local Larmor
frequency $\f \w_{\mathrm{L}}(\f r)$ that we assume to be parallel to $\z$. To that
end we substitute $\f S(t)$ with $\f\R(t)\f S(t)$ and similarly for $\f
J_i$, where $\f\R=\f\R(-\f\w_{\mathrm{L}} t)$ is a  3D rotation matrix. Then we get from~(\ref{s}), (\ref{j})
\bea
\p_t{\f S} &+& \widetilde D_i\f J_i=0,\label{LarS}\\
\p_t{\f J_i} &+& \F{w^2}3\widetilde D_i(\f S-\f\w_{\mathrm{L}}) + \kappa\f S\X\f
J_i=-\F{\f J_i}{\tau_1},\label{LarJ}
\eea
where $\widetilde D_i=(\nabla_i+g\widetilde{\f A}_i\X)$ and $\widetilde{\f A}_i=\f\R^{-1}\f A_i$.

We here used the fact that $e_{\A ij}\f\R_{i\B}\f\R_{j\G}=\f\R_{\A k}e_{k\B\G}$,
which follows from the general equality holding for an
arbitrary matrix: $e_{\A ij}\f\R_{\A k}\f\R_{i\B}\f\R_{j\G}=e_{k\B\G}\det\f\R$.

Eq.~(\ref{LarJ}) is an inhomogeneous linear differential equation on ${\f J_i}$.
Its solution can be represented as the sum of the general solution of the
corresponding homogeneous equation and a particular solution of
the inhomogeneous one. The general solution of the homogeneous equation contains the
factor $\exp(-t/\tau_1)$ and dies out in a time $\tau_1$. Thus for the most of the time of an experiment we may take
into account only the  particular-solution part of the spin current.

In the absence of electric fields $\widetilde D_i=\nabla_i$ and both the inhomogeneous term and
the coefficients in  Eq.~(\ref{LarJ}) are independent of time. In this case Eqs. (\ref{LarS}), (\ref{LarJ}) allow
exactly stationary solutions, wherein $\p_t{\f S}$ and $\p_t{\f J_i}$ are exactly zero in the Larmor frame.

 In the presence of an electric field, the inhomogeneous term in (\ref{LarJ})
 contains, besides the part constant in time, a fast contribution from the rotation matrix that oscillates with the Larmor
 frequency $\w_{\mathrm{L}}$. Hence the solution will be the sum of  constant and  oscillating terms. The oscillating term
 in the Larmor frame will appear from
 the point of view of the laboratory frame as a doubled frequency motion and must induce a
 doubled frequency harmonic in the NMR signal.

 However the amplitude of this harmonic as well as the correction to the stationary motion due to electric
 field will be of the order of smallness of $gEL/c$, where $L$ is the spatial scale of the problem. Therefore we can
 expand $\f S$ and $\f J_i$ into degrees of electric field to within the first order: $\f S(\f E)=\f S+\D\f S$,
and similarly for $\f J_i$, where $\f S$ and $\f J_i$ now designate the corresponding quantities in the absence of
electric field that meet the conventional Leggett equations. For the departures $\D\f S$ and $\D\f J_i$
proportional to the electric field we obtain from~(\ref{LarS}), (\ref{LarJ})
\bea
\p_t{\D\f S} &+& \nabla_i\D\f J_i+g\widetilde{\f A}_i\X\f J_i=0,\label{LarDS}\\
\p_t{\D\f J_i} &+& \F{w^2}3(\nabla_i \D\f S +g\widetilde{\f A}_i\X(\f S-\f\w_{\mathrm{L}}))+ \kappa(\D\f S\X\f
J_i+\f S\X\D\f J_i)=-\F{\D\f J_i}{\tau_1}.\label{LarDJ}
\eea

 Thus $\D\f S$ and $\D\f J_i$ (deemed as components of one unknown quantity) satisfy a system of inhomogeneous linear
 differential equations with coefficients depending on coordinates but independent of time. The electric field
 enters the equations only through inhomogeneous terms and hence the corresponding homogeneous system describes the time
 evolution of small perturbations to a known unperturbed zero-electric-field solution. The general solution of the
 homogeneous system is the sum of
 particular solutions in which  $\D\f S$ and $\D\f J_i$ depend on time through the factors $e^{-i\w t}$. The perturbation
 frequencies $\w$ are to be determined by solution of the homogeneous system with the corresponding boundary conditions.
 For the unperturbed distribution to be stable the imaginary parts of all the possible frequencies $\w$ need to be
 negative. Then the arisen perturbations will die out exponentially.

 To find the quasistationary part of the solution (i.e. the one that is constant in time in the Larmor frame), we should
 average the  inhomogeneous terms     over the period of precession $1/\w_{\mathrm{L}}$, which yields $\langle \widetilde{\f
 A}_i\rangle=A_i^z \z$.

We remind that $A_i^z=-c^{-1}e_{zik}E_k$. Thus the $\z$-component of the electric field have no effect on the
quasistationary dynamics. If we choose $\f E\parallel\h{\f x}$ for definiteness, then $A_i^z=c^{-1}E\h{\f y}_i$ and the
electric field falls out from the equations for the $\h{\f  x}_i$- and $\h{\f  z}_i$- orbital components of the spin current, i.e. the
electric field affects the quasistationary motion only if the $\h{\f  y}_i$-components of the spin current is non-zero.

\section{One-Dimensional Flow}

As an example we consider spin flow through a thin channel directed along
the $\h{\f  y}_i$-axis.  But first we  examine the spin distribution in such a geometry in the absence of an electric field.
The situation is reminiscent of the flow along the $\h{\f  z}_i$-directed  channel studied experimentally~\cite{Nunes} and
theoretically~\cite{Ragan}. We may extend  the results obtained in Ref.~\cite{Ragan} to our situation since
in the derivation of distributions and in the stability analysis in Ref.~\cite{Ragan} only the fact
that the flow is one-dimensional is used and the name of the coordinate along which the distribution changes makes no difference.

\subsection{In the Absence of  Electric Field}

\begin{figure}
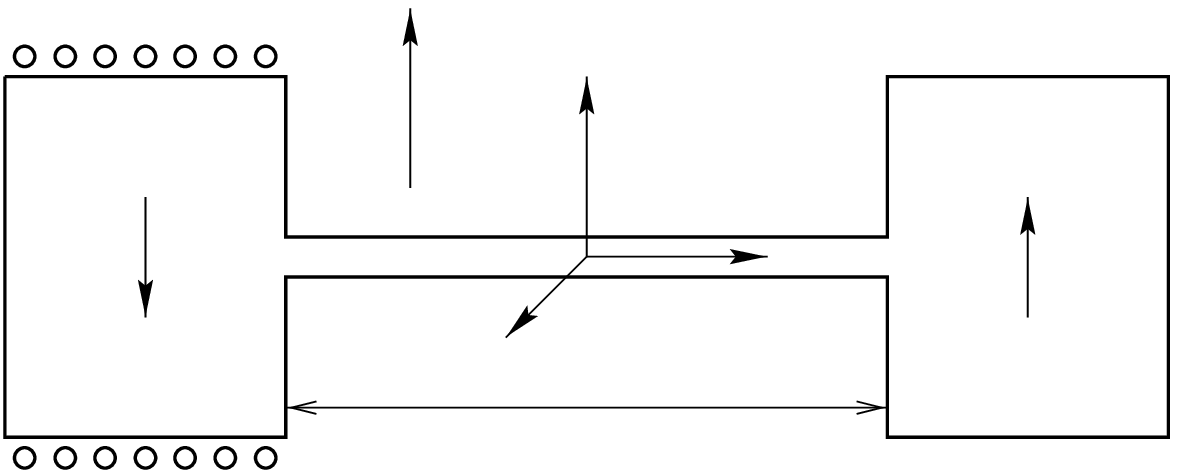
\vspace{.5cm}
\caption{Geometry of idealized spin-diffusion experiment. The circles are turns of wire of the rf coil used to invert
the magnetization in the left reservoir.}\label{kanal}
\end{figure}

In the idealized geometry of Ref.~\cite{Ragan} two reservoirs of a Fermi-liquid polarized by an external magnetic field
directed along $\z$ are
connected via a long thin tube of length $2L$ and cross-sectional area $A\ll L^2$. After a $\pi$-pulse of a rf magnetic
field, inverting the spins, is applied to one of the volumes, diffusion tends to diminish the
large longitudinal polarization gradient along the channel and eventually equalize polarization in the two chambers (see
Fig.~\ref{kanal}).

Since the time $L^2/D$ to diffuse through the tube is much longer than the time $A/D$ to diffuse away from the tube
entrances, we may, following~\cite{Ragan}, assume fixed boundary conditions $\f S(y=\pm L)=\pm S_0\z$, where $S_0$ is
the magnetization density
in the reservoirs. For convenience we transform to the dimensionless units: $s=y/L$, $\tau=D_0 t/L^2$, $\f m=\f S/S_0$, $\f
j=L\f J_y/D_0 S_0$, and introduce the parameter $v=-\kappa S_0\tau_1$. Here $D_0=w^2\tau_1/3$ is the diffusion
coefficient. As a matter of fact, $\kappa$ is positive for
real liquids and thus $v$ is negative.

Now we rederive the results of~\cite{Ragan} generalized to  the  case  of
the parameter $\h{\f m}_0\ne \h{\f x}$ (see below).

The Leggett equations in the Larmor  frame, which are just eqs.~(\ref{LarS}),
(\ref{LarJ}) with $\widetilde D_i=\nabla_i$, for stationary solutions
depending on only one spatial coordinate yield
\be
\p_s\f j=0,\qquad(\p_s+v\f j\X)\f m +\f j=0.
\ee
We get
\be
\f j=\const, \qquad\f m(s)=\f\R(-sv\f j)\f m_0-s\f j,
\ee
where $\f m_0\equiv\f m(s=0)$ is the magnetization in the center of the channel.
Its absolute value is $\sqrt{1-j^2}$ and its direction remains an arbitrary
parameter. This uncertainty in the direction of $\h{\f m}_0$
corresponds to the indefiniteness of the initial phase of the
precession. Actually the result in~\cite{Ragan} is written out (with some slips) in the coordinate representation corresponding to
the choice of $\f m_0\parallel\h{\f x}$. This is why we dwelt on the derivation.

Then the boundary conditions $\I.\f m\J|_{s=\pm 1}=\pm \z$
require $\f\R(2v\f j)\f m_0=-\f m_0$.
Hence either $\f m_0=0$ and we get
\be
\f j=-\z, \qquad\f m(s)=s\z.\label{solution1}
\ee

Or $\f j\perp\f m_0\ne 0$ and $vj=\pi/2+\pi n$,  $n\in\mathsf{Z}$. The stability analysis and numerical simulations~\cite{Ragan} show that only the solutions with $n=0$ and
$n=-1$ are stable. For liquids with $v$ negative $n=-1$ is relevant, while $n=0$ corresponds to $v$ positive.
Using the overt form of the rotation
matrix $\f\R(\f\theta)=(\D_{\A\B}-\h\theta_\A\h\theta_\B)\cos\theta+
\h\theta_\A\h\theta_\B-e_{\A\B\G}\h\theta_\G\sin\theta$,
we obtain for $n=-1$:
\be
j=-\F{\pi}{2v},\quad\h{\f\jmath}=-j\z+\f m_0\X\z,
\quad\f m(s)=-\f m_0\X\h{\f\jmath}\sin(\pi s/2)+
\f m_0\cos(\pi s/2)-sj\h{\f\jmath}.\label{solution}
\ee
In such a solution the polar angle $\pi-\vartheta$ between
$\h{\f\jmath}$ and $\z$ is fixed by the condition:
$j=-(\h{\f\jmath}\z)=\cos\vartheta$. (Here we see that of
necessity $j\le 1$ and $|v|\ge \pi/2$.)
The arbitrary azimuthal angle
of $\h{\f\jmath}$ is
parameterized by a constant unit vector $\h{\f m_0}=\z\X\h{\f\jmath}/m_0$.

\begin{figure}
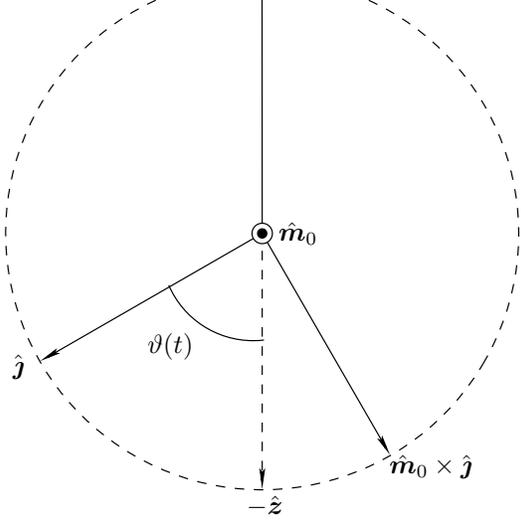
\vspace{.5cm}
\caption{Disposition of the right-handed basis in the non-longitudinal
one-dimensional
stable stationary solution of Leggett's equations with the
boundary conditions as on Fig. 1.
This solution is realized when
$|v|\ge\pi/2$. The unit vector $\h{\f m}_0$ is directed from
the plane of the picture
(the $\h{\f\jmath}-\z$ plane) towards
the reader.
 The angle $\vartheta$ equals $\vartheta=\arccos(\pi/2|v|)$.
It varies from $\vartheta(0)\in(0,\pi/2)$ at the start to zero at the
time of switching to the longitudinal solution $t=t^*-T/2$.
}\label{circ1}
\end{figure}

The three unit vectors $\h{\f
m_0}\X\h{\f\jmath}$, $\h{\f m}_0$, $\h{\f\jmath}$ form a right-handed basis (see Fig.~\ref{circ1}).

Subject to the value of $|v|$ either of the above stable stationary solutions
takes place. For $|v|\le\pi/2$ the solution is the longitudinal
diffusion~(\ref{solution1}).
For $|v|\ge \pi/2$ this configuration becomes unstable
(Castaing instability) and~(\ref{solution}) is realized.

Properly speaking, arbitrary boundary conditions can be considered in the same manner but only the ones involved, i.e.
of the diametrically opposed magnetizations in the two reservoirs, appear to have a solution of more or less simple form.

In fact, the diffusion under consideration is only \emph{ quasi}stationary.
That is, it is independent of time on the short scale $L^2/D_0$. But on
long times
owing to the finite size of the reservoirs, slowly as it
will, the current will change the magnetization densities in the reservoirs, thus changing the boundary conditions.
Thanks to this slowness we may deem the situation  at any given moment
as stationary \emph{on times} $\sim\tau$ and simply
substitute $\hS_0$ for $\z$ in the boundary conditions, so that now they read $\f m(s=\pm 1)=\pm\hS_0$.

For the long-time dependence of the magnetization in the right reservoir we then get
\be
T\F{d\f S_0}{dt}=S_0\f j,\label{s_0}
\ee
where $T=VL/D_0A$ is the long-time scale. Thus the quasistationary
approximation is valid as long as $T\gg L^2/D_0$, or, equivalently, $V\gg LA$. For experimental conditions
of~\cite{Nunes} $T\sim 100$ sec.

For $|v|\le\pi/2$ the current $\f j=-\hS_0$ and the
magnetization decays exponentially:
\be
\f S_0(t)=\f S_0(0)\exp(-t/T).
\ee

For $|v|\ge\pi/2$ the magnetization density diminishes algebraically  until it reaches the critical value $\pi/2$. It is
convenient to introduce a constant $\zeta=\pi/2\kappa\tau_1$ so that $j=\zeta/S_0$.
Upon substituting $\f j=j(-j\hS_0+\sqrt{1-j^2}\h{\f m}_0\X\hS_0)$ eq.~(\ref{s_0}) yields
\be
T\F{d S_0}{d t}=-\F{\zeta^2}{S_0},\qquad T\F{d\hS_0}{dt}=\F{\zeta}{S_0}\sqrt{1-\F{\zeta^2}{S_0^2}}\h{\f m}_0\X\hS_0.
\ee

Then while
$v(t)$ exceeds the critical value, i.e. for times $t\le t^*-T/2$, where $t^*=TS_0^2(0)/2\zeta^2$,  we get
\be
S_0(t)=\zeta\sqrt{2\F{t^*-t}{T}};
\qquad\hS_0(t)=\f\R\I(\h{\f m}_0\int\limits_0^t
\F{\zeta}{S_0}\sqrt{1-\F{\zeta^2}{S_0^2}}\F{dt}T\J)=\f\R(\h{\f m}_0\B(t))\z,
\ee
where
\be
\B(t)=\I[\vartheta(t)-\sqrt{2\F{t^*-t}{T}-1}\J]_0^t
\approx j(0)\sqrt{1-j^2(0)}\;\F tT+O(\F{t^2}{T^2}).\label{B}
\ee

\begin{figure}
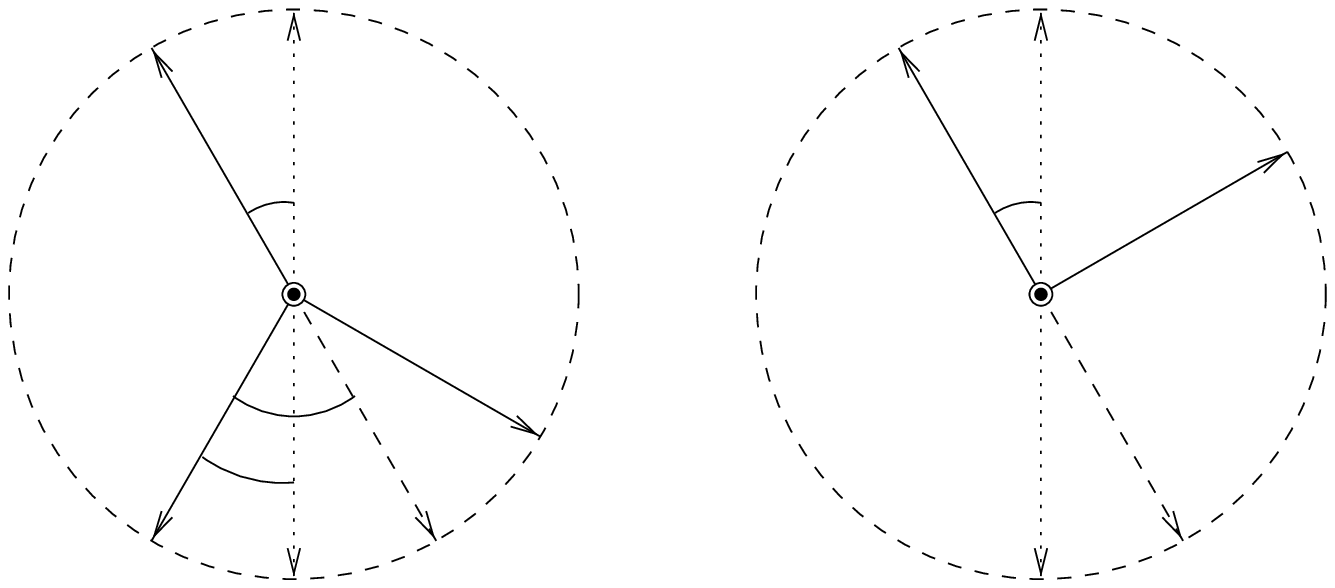
\vspace{.5cm}
\caption{Disposition of the right-handed basis:
$a)$ for the non-longitudinal
solution at arbitrary non-zero times and $b)$ for the longitudinal one following the
non-longitudinal.
The angle $\B$ becomes non-zero during diffusion, at the start $\hS_0\equiv\z$.
}\label{circ}
\end{figure}

The absolute value of magnetization decays as a square root of time and
its direction rotates around $\h{\f m_0}$, i.e. in the $\h{\f
\jmath}-\z$ plane (see Fig.~(\ref{circ}a)). The angle $\B$ increases monotonically with time. As numerical evaluation
shows, it  reaches $\pi$ at the critical point for the first time when the duration of the non-longitudinal part of diffusion
constitutes $t^*\approx10 T$.

Note that the
time evolution of the magnetizations in the two reservoirs is symmetric as a result of the conservation of current. And at
any moment the directions of the magnetizations in the vessels remain opposite to ensure the diametrically opposed
conditions involved.

\begin{figure}
\sf
\input{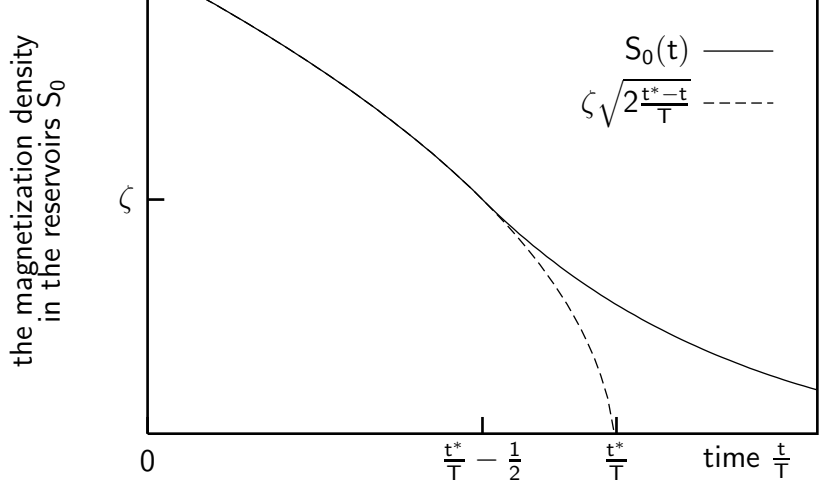}
\normalfont
\caption{ The time dependence of the absolute value of the magnetization density in the reservoirs $S_0(t)$ in case the
initial value $S_0(t=0)$ exceeds the critical value $\pi/2$ at which the longitudinal diffusion becomes unstable. At
$t=t^*-T/2$  the value of $S_0$ reaches $\pi/2$ and from this point  diffusion switches to the longitudinal regime.
}\label{S_0}
\end{figure}

For $t\ge t^*-T/2$ the magnetization will proceed by exponential slump
(see Fig.~(\ref{S_0})) with the angle $\B_f=\B(t^*-T/2)\ne 0$
between $\z$ and
$\hS_0$ (see Fig.~(\ref{circ}b)).

\subsection{In the Presence of an Electric Field}

Let now an electric field $\f E=E\h{\f x} $ be applied to the liquid in
the channel at some time $t_E>0$. In accordance with what was said above,
all quantities will acquaint increments proportional to the electric field:
the magnetization and current in the channel $\f m\to\f m+\D\f m$, $\f j\to
\f j+\D\f j$. And the magnetization in the right reservoir $\f S_0\to\f S_0+
\D\f S_0$.

The boundary condition for $\D\f m$ will be $\D\f m(s=\pm 1)=\pm \D\f S_0/S_0$.

Since the quasistationary inhomogeneous terms $(g\langle \widetilde{\f A}_i\rangle\X)$ are independent of time, we
 seek for a quasistationary solution $\p_\tau{\D\f m}=0$, $\p_\tau{\D\f j}=0$. The oscillating terms will be discussed later.
Of course, $\f S_0$, $\f m$ and $\f j$ all depend on the ``long'' time $t/T$.

 After introducing the  additional parameter $\e=gEL/c$ we get
\bea
\p_s\D\f j &+&\e\z\X\f j=0,\label{dj}\\
(\p_s+v\f j\X)\D\f m&+&\e\z\X\f m+\D\f j+v\D\f j\X\f m=0.\label{dm}
\eea

The first of the equations yields
\be
\D\f j(s) =\D\f j_{\e}(s)+\D\f j_0,\label{dj(s)}
\ee
 where
\be
\D\f j_{\e}(s)= -\e\z\X\f j s\label{dje}
\ee
 and $\D\f j_0$ is a constant to be
determined from  eq.~(\ref{dm}) with the boundary conditions
 imposed. We can make a substitution
$\D\f m(s)=\f \R(-vs\f j)\D\tilde{\f m}(s)$ into  eq.~(\ref{dm}), after
which it can be easily integrated
\be
\D\f m(s)=\R(-vs\f j)\int_{-1}^s\f \R(vs\f j)(\e\z\X\f m+\D\f j+v\D\f j\X\f m) ds-\F{\D\f S_0}{S_0}.
\ee

And then the
boundary conditions reduce to
\be
\int_{-1}^1\f \R(vs\f j)(\e\z\X\f m+\D\f j+v\D\f j\X\f m) ds=2\R(v\f j)\F{\D\f S_0}{S_0}.\label{fix}
\ee
As a matter of fact, $\D\f j_0$ should be determined from this equality.
At the moment $t_E$ of switching on of the electric field $\D\f S_0=0$
and the above equation can be solved analytically.
$\D\f j_0$ then turns out to be zero.
Nonetheless, although $\D\f j_0$ is, generally speaking, non-zero
at arbitrary times, this part of $\D\f j$ is
constant in space and does not change the symmetric picture of the diffusion.

It is the part $\D\f j_{\e}$ of $\D\f j$ that depends on $s$ and brings about an observable effect of interest to us. This part has
opposite signs at the two tube entrances, thus changing the magnetizations in the two reservoirs by the same amount, i.e.
\emph{not symmetrically}.

Unless $\f j$ is parallel to $\z$, the term $\D\f j_{\e}$ is directed along $\h{\f m}_0$ perpendicular to the
$\h{\f \jmath}-\z$ plane. Therefore, as diffusion goes on, the non-symmetric change of magnetization in the reservoirs
will lead to an \emph{ additional phase shift} between the two vessels.

To find the change in the time evolution of $\f S_0$ due to electric field we substitute $\f S_0$ with $\f S_0+\D\f S_0$,
where $\f S_0(t)$, as previously, belongs to the case of the absence of electric
field and the part $\D\f S_0$ is proportional to the electric field. This part  obeys an equation, which can be obtained by taking
variations of eq.~(\ref{s_0}) using (\ref{dj(s)}):
\be
T\F{d\D\f S_0}{dt}=S_0\I[\D\f j_0+\D\f j_{\e}(s=1)\J]+\D S_0\f j.
\ee

Next we expand $\D\f S_0$ onto the components in and perpendicular to the $\h{\f \jmath}-\z$ plane: $\D\f S_0=\D\f
S_0^{m}+\D\f S_0^{j-z}$. For the component $\D\f S_0^{m}$ perpendicular to the $\h{\f \jmath}-\z$ plane we get
$T{d\D S_0^m}/{dt}=S_0[\D j_0^m+\D j_{\e}(1)]$. The phase \emph{shift} is
described by the term $\D j_{\e}$. Hence the phase shift  increases with
time as
\be
\D\A(t)\approx-2\F{\D S_0^m(t)}{ S_0(t)\sin\B(t)}=-2\F{\int_{t_E}^t
S_0(t)\D j_{\e}(1;t)\F{dt}{T}}{ S_0(t)\sin\B(t)}\label{da},
\ee
where we have neglected the presumably small component $\D \f S_0^{j-z}$ compared to $\f S_0$ in the denominator.

For the non-longitudinal stable solution~(\ref{solution}) we have $\D j_{\e}(1)=\e j\sin[\B(t)-\vartheta(t)]$.
and the corresponding integral can be found only numerically. For
times close to $t_E$ we have ($t>t_E$)
\be
\D\A\approx 2\e\F{t-t_E}{T}j(t_E)\F{\sin[\B(t_E)-\vartheta(t_E)]}{\sin\B(t_E)}.
\ee

When $t_E\ll T$, the last expression simplifies
\be
\D\A\approx -2\e\F{t-t_E}{t_E}.
\ee
However, this formula cannot be applied to $t_E$ extremely close to zero.
Namely, in the case that the transverse component of the magnetization in the
reservoir $S_0(t_E)\sin\B(t_E)$ is of the order of magnitude of $\D S_0^m$,
the up-to-the-first-order extension of the transverse component of $\f S_0$
in electric field is not valid.

For  the longitudinal stable solution~(\ref{solution1}) eq.~(\ref{da})
can be easily integrated and yields $\sin\B_f$ in both the numerator
and the denominator, where $\B_f\equiv\B(t^*-T/2)$ here is the  polar angle between $\hS_0$
and $\z$ handed down from the first
 (non-longitudinal) part of the diffusion. Thus
\be
\D\A(t)=-2\e[\exp((t-t_0)/T)-1]\label{long}
\ee
for all values of $\B_f\equiv\B(t^*-T/2)$ except for $\B_f=\pi n$, $n\in\sf Z$,
when the phase shift is undefined. In particular, if there were no
non-longitudinal regime at all (due to insufficient initial polarization),
and hence the magnetization is parallel to $\z$, there would be no additional
phase shift.

The time $t_0$ in~(\ref{long}) is either $t^*-T/2$, when the longitudinal
solution follows the non-longitudinal regime, or $t_E$ if an electric field
is turned on already at the longitudinal part of the diffusion.
In the former case,
eq.~(\ref{long}) should be added up with the phase shift accumulated during
the non-longitudinal part.

 Exp.~(\ref{long}) formally tends to infinity with time, but owing to the
 small coefficient $\e$ it still remains very small for all of the time of an experiment.

The appearance of an additional phase shift in an electric field is quite
understandable. The diffusion process may be outlined as a transfer of
down spins from the left reservoir to the right one. Naturally, the
current along $\h{\f y}_i$ of down spins is equivalent to the current of up
spins in the opposite direction. In accordance with the boundary
conditions the spins of $^3$He atoms at the left end of the tube are
directed preferably down. The ``surplus'' of down spins drifts \emph{ to the
right} and interacts with the electric field $\f E=E\h{\f x}$ as if there was
an \emph{ increase} to the magnetic field: $-(Ev_{\mathrm{F}}/c)\f\mu\z$.

On the contrary, the up spins at the right end float \emph{ to the left},
and their spin-orbital energy $(Ev_{\mathrm{F}}/c)\f\mu\z$ amounts to an effective
\emph{ decrease} of the magnetic field. The spins at the opposite ends thus
precess at slightly different frequencies, unless, of course, they are
not confined to the $\z$ direction like in the purely longitudinal
solution. Hence in the second solution (for $|v|\ge\pi/2$) there develops
a phase shift between the two entrances of the tube.

  The conclusion about the existence of an additional phase shift was drawn for a situation with diametrically
opposed magnetization directions at the tube ends as the boundary condition. But the inference in itself remains valid
for other boundary conditions as well, irrespective of the fact that we are ignorant of
zero-electric-field distribution in that case and of exact expression for the phase shift. This follows, first, from the
current conservation in the absence of electric field, which secures the self-similar boundary conditions in the
quasistationary diffusion process. And secondly, from
eq.~(\ref{dj(s)}), which holds for all stationary solutions in one-dimensional geometry and leads to non-symmetric change
of the magnetization in the reservoirs when an electric field is applied.

As for the oscillating part of the gauge field, it does not
contribute to the current along the channel. Indeed, the oscillating part of the gauge field is
\be
g\widetilde{\f A}_i=(gE/c)\z_i(\h{\f x}\sin \w_{\mathrm{L}} t-\h{\f y}\cos\w_{\mathrm{L}}t)
\ee
 if the
electric field is parallel to $\h{\f x}_i$ as before, that is, $g\widetilde{\f A}_i$ is proportional to $\z_i$. Taking into account that different
$i$-components of $\D{\f J}_i$ do not enter each other's evolution equation~(\ref{LarDJ}), it means that in the particular solution of
the inhomogeneous system only $\D\f S$ and $\D\f J_z$ will be non-zero and $\D\f J_x$ and $\D\f J_y$ will figure only in
the general solution of the homogeneous equation. According to what was said above, if only the unperturbed solution is
stable, the latter dies out exponentially with the rate $(\Im \w)^{-1}$, where $\w$ is an eigenvalue
of~(\ref{LarDS}),(\ref{LarDJ}).

To be exact, the oscillating part of the gauge field brings about the
appearance of a fine structure across the channel, which
vary on a spatial scale of $\sqrt A$ and oscillates with the Larmor frequency in the Larmor frame. Though in
general it is rather complicated, the main features of such a structure may be seen on the easier example of the
longitudinal diffusion, which is considered in the Appendix.

\section{Strongly-Polarized Fermi-Liquid}
Landau theory cannot be literally applied to  strongly spin-polarized Fermi liquids~\cite{Fom}.  However for $T=0$
Fomin derived microscopically transverse spin dynamics equations~\cite{Fom}, the form of which, in fact,
coincides with the collisionless limit of the Leggett equations.

In this section we will consider the application of the SU(2) gauge invariance
directly to a microscopic Hamiltonian of a liquid at $T=0$. The Hamiltonian of the liquid involved
in the absence of external fields consists of the kinetic energy\footnote{We leave out spin
indices of the field operators of $^3$He atoms $\psi$ for brevity.}
\be
\Ham_{\mathrm{K}}=-\F1{2m}\int \psi^+(\f r)\nabla_i^2  \psi(\f r) \;d^3 \f r
\ee
and an interaction $\Ham_{\mathrm{int}}$ between  particles. We neglect the very small spin-orbital interaction \emph{
between particles}.

To include electric and magnetic fields into consideration we must use~(\ref{Ham}) instead of
$\Ham_{\mathrm{K}}$. Such a replacement can be interpreted as a result of a gauge transformation
$\p_\mu\to D_\mu=\p_\mu-ig\pau\f A_\mu$. It is convenient to write
the Hamiltonian~(\ref{Ham}) in the form:
\be
\D\Ham=-\int g\f A_0(\f r)\f S(\f r)\;d^3 \f r-\int g\f A_i(\f r)\f J_i(-g\f
A_i,\f r)\;d^3 \f r,
\ee
where the spin and the spin-current operators are defined as follows
\bea
\f S(\f r)&=& \psi^+(\f r)\pau\psi(\f r),\\
\f J_i(\f X_i,\f r)&=&\F1{2m}  \I[\I(i\nabla_i\psi^+(\f
r)\J)\pau \psi(\f r)+ \psi^+(\f r)\pau \I(-i\nabla_i\psi(\f
r)\J)+\F14\psi^+(\f
r)\psi(\f r) \f X_i
\J].
\eea

Following Fomin~\cite{Fom} we now transform in each space point to a
frame precessing at such a frequency $\f\Omega$, that the
magnetization in the rotating frame  is equilibrium. $\f\Omega$ is supposed to vary slowly enough in space
and time so that we could leave only first terms in $\w$ and $ k$, where $\w$
and $k$ are respectively the characteristic frequency and wave number of
$\f\Omega$ in the Larmor reference system.

As a result we get the following expression for the Hamiltonian in the
rotating frame
\be
\Ham=\Ham_{\mathrm{K}}+\Ham_{\mathrm{int}}+\delta\Ham+\int \f S(\f r)\f\Omega\;d^3 \f r.
\ee

Now we introduce $\A$ and $\B$ as the spherical coordinates of
$-\f\Omega$: $\f\Omega=-\Omega(\sin\B\cos\A, \sin\B\sin\A, \cos\B)$ and rotate the spin space
$\psi\to\R\psi$, where
\be
\R=\exp(-i\A\sigma^z/2)\exp(-i\B\sigma^y/2)\exp(-i\G\sigma^z/2)
\ee
is a finite rotations operator expressed via Eulerian angles. The angle
$\G$ is a free parameter. The handy constraint on it will be clear later.
Then the space-time derivatives  $\p_\mu\psi$ transform as follows:
\be
\p_\mu\psi\to\R\p_\mu\psi+(\p_\mu\R)\psi=\R D_\mu\psi,
\ee
 with  the covariant derivative
$D_\mu\psi=\p_\mu\psi+(\R^+\p_\mu\R)\psi=\p_\mu\psi-i\pau\f\w_\mu\psi$.
This must be regarded as the definition of $\f\w_\mu$. It can be written out equivalently as
\be
\f\w_\mu=\F12e_{\A\B\G}\f\R_{\B j}\p_\mu \f\R_{\G j}
= \I(\ba{rcr} -\sin\B\cos\G\,\p_\mu\A & + & \sin\G\,\p_\mu\B \\
\sin\B\sin\G\,\p_\mu\A & +
& \cos\G\,\p_\mu\B \\
\cos\B\,\p_\mu\A & + & \p_\mu\G \ea\J),
\ee
where $\f\R_{\A i}=\f\R_z(-\G)\f\R_y(-\B)\f\R_z(-\A)$ is a
3D rotation matrix.

Under this transformation
\bea
\Ham&\to&\Ham_{\mathrm{K}}+\Ham_{\mathrm{int}}+\int \f S(\f r) \f\R\f\Omega
\;d^3 \f r+\I.\D\Ham \J|_{g\f A_\mu\to g \f\R\f A_\mu+\f\w_\mu}=\nonumber\\
&=&\Ham_{\mathrm{K}}+\Ham_{\mathrm{int}}+\int\f S(\f r) \f\R\f\Omega \;d^3 \f r
-\int(\f\w_0+\w_{\mathrm{L}} \f\R\z)\f S(\f r)\;d^3 \f r-\nonumber\\
&-&\int(\f\w_i+g\f\R\f A_i)\f J_i(-\f\w_i-g\f\R\f A_i,\f r)\;d^3\f r.\label{base}
\eea

As can be easily seen,
$\f\R\z=(-\sin\B\cos\G, \sin\B\sin\G, \cos\B)$,
$\f\R\f\Omega=-\Omega\z$.

In case that the absolute value of $\f\Omega$ is greater than its variations in the reference system that precesses at the
Larmor frequency (i.e. greater than $\dot\A+\w_{\mathrm{L}}$, $\dot\B$, $\dot\G$ and
$\nabla_i^2\A/m$,
$\nabla_i^2\B/m$, $\nabla_i^2\G/m$), then the first three terms in (\ref{base}) are the
greatest. Together they constitute the unperturbed Hamiltonian of a Fermi liquid in a uniform field
$-\Omega\z$. The ground state of this Hamiltonian is $\langle\f
S\rangle=(0,0,S)$ and $\langle\f J_i\rangle=0$. The former is extremely important because shows that
$\A$ and  $\B$ serve as \emph{ spherical coordinates of $\f S$} in the \emph{ laboratory} frame.

The other terms in (\ref{base}) are the perturbation. Its magnitude in the ground state equals
\be
-(\f\w_0+\w_{\mathrm{L}}\f\R\z)\langle\f
S\rangle=-S(\dot\G+\cos\B(\dot\A+\w_{\mathrm{L}})).
\ee

The expression above contains only time derivatives. To include space
derivatives one should allow for the second order corrections, which
can
be evaluated as in Maki's paper~\cite{Maki}:
\bea
\langle\Delta\Ham\rangle&=&\Delta
F=-\F{\chi^J}2(\f\w_i+g\f\R\f A_i)^2\nonumber\\&=&-\F{\chi^J}2
[(\f\n\A)^2+(\f\n\B)^2+(\f\n\G)^2+2\cos\B\f\n\A\f\n\G]-\chi^J \F{g^2}{c^2}E^2+\chi^Je_{\A
ik}\tilde\w_{\A i}\F{g}{c}E_k.
\eea

Note  the wrong sign in Ref.~\cite{Maki}. A  new definition is
\be
\tilde{\f\w}_{ i}=\f\R^{-1}\f\w_i= \I(\ba{rcr} \sin\B\cos\A\,\nabla_i\G & -
&
\sin\A\,\nabla_i\B \\ \sin\B\sin\A\,\nabla_i\G & +
& \cos\A\,\nabla_i\B \\
\cos\B\,\nabla_i\G & + & \nabla_i\A \ea\J).
\ee

Thus the effective Lagrangian ${\cal
L}_{\mathrm{eff}}=-S(\dot\G+\cos\B(\dot\A+\w_{\mathrm{L}}))+\langle\Delta\Ham\rangle$.
Varying this with respect to $\G$, we obtain
\be
\dot S+\chi^J\f\n(
\f\n\G+\cos\B\f\n\A+\hS\X\F{g}{c}\f E)=0.\label{G}
\ee

In the end we could set a constraint on $\G$ guided by the reasons of
rationality, i.e. so
that $\dot S=0$, which requires $\f\n\G+\cos\B\f\n\A=-\hS\X g\f
E/c$. Substituting this condition into
the equations obtained by varying ${\cal L}_{\mathrm{eff}}$ with respect to $\A$
and $\B$, we obtain a set of two equations for $\A$ and $\B$ which are equivalent to
\be
D_t\hS=\F{\chi^J}{S}D_i(D_i\hS\X\hS). \label{strong}
\ee

It is interesting to trace the \emph{ formal} transition from Eqs.~(\ref{s}), (\ref{j}) to Eq.~(\ref{strong}) though the
modified Leggett equations, definitely, cannot be applied to a strongly-polarized liquid.

First of all, for a strongly-polarized liquid $S$ is much greater than $\w_{\mathrm{L}}$, hence, $\f S^{\mathrm{eq}}$ could be neglected
in~(\ref{j}). Moreover, the collision integral $\f J_i/\tau_1$ vanishes as compared to the molecular-field term
$\kappa\f S\X\f J_i$ by virtue of the parameter $\kappa S\tau_1\gg 1$. Thus we can get to the limit $\kappa S\tau_1\gg
1$ in an expression for the quasistationary current, which can be obtained by resolving Eq.~(\ref{j}) with respect to
$\f J_i$ provided that $D_t\f J_i=0$:
\be
\f J_i=-\F{D_0}{1+(\kappa S\tau_1)^2}[D_i\f S+\kappa\tau_1\f S\X D_i\f S+(\kappa\tau_1)^2\f S( S\nabla_i
S)].\label{statJ}
\ee

As was discussed concerning the weak-polarization case, the transition to the quasistationary current is self-consistent only for
the quasistationary part of the solution because only then both sides of~(\ref{statJ}) are constant in time in the
Larmor frame.
The senior (the third) term in~(\ref{statJ}) identically vanishes for homogeneous spatial
distributions of the absolute value of magnetization. We assume this condition to fulfil, for the collisionless regime
of the weak-polarization case it can be proved.

Substituting then the expression for the quasistationary current into~(\ref{s}), we obtain~(\ref{strong}) with
$\chi_J=w^2/3\kappa$, or, in ordinary units, $\chi_J=(\chi_{\mathrm{n}}/g^2)(w^2/3\kappa)$. It is this value of $\chi_J$ that the
Fermi-liquid theory gives in the limit of weak polarizations~\cite{Fom}. Still $\chi_J$ cannot be obtained in the
general case.

From the fact that the modified Leggett equations~(\ref{s}), (\ref{j}) yield~(\ref{strong}) in the deep collisionless
limit in the quasistationary approximation it follows immediately that all conclusions about the quasistationary part of
spin flow through a channel and, in particular, of the additional phase shift remain valid in the strong-polarization case.

\section{Conclusion}

In this paper we considered how an external electric field can influence spin dynamics of an electrically neutral Fermi
liquid through the spin-orbital interaction with nuclear magnetic moments. In the framework of Landau's theory of Fermi
liquids attributing an additional energy due to spin-orbital interaction to each quasiparticle leads to generalization of
Leggett's equations. The amendments are consistent with the requirements imposed by intrinsic SU(2) gauge invariance of
the interaction of a magnetic moment with electromagnetic field.
The corrections caused by electric field in the
Leggett's equations are responsible for an additional phase shift in a
one-dimensional spin flow experiment. This phase
shift, which is proportional to the electric field, coincides in the order
of magnitude with that  in superfluid
$^3$He-$B$ but grows with time. Its magnitude is observable experimentally.

For strongly spin-polarized Fermi liquids, electric field is incorporated
into Fomin's equations for transverse spin dynamics of a strongly spin-polarized Fermi liquid at zero temperature. The
formal correspondence between these equations and the collisionless limit of Leggett's equations spreads to the case of
an electric field present as well.

\section*{Acknowledgments}
I appreciate the scientific supervision of V.~P.~Mineev, who
raised the task, introduced me to the problem and without whom this paper would never appear. I am also thankful to
V.~V.~Dmitriev and I.~A.~Fomin for discussions.  This work
was performed with partial financial support from the Statistical Physics Program of the Ministry of Sciences of the
Russian Federation, by Grants Nos. 96-02-16041b and 96-15-96632 from the Russian Fund for Bare Research (State Program
for Support of Scientific Seminars), by the Landau Scholarship from Forschungszentrum J\"ulich  and from the INTAS Program (Grant 96-0610).

\section*{Appendix}

In the case of longitudinal diffusion the inhomogeneous term in  (\ref{LarDS})  is identically
zero, but not in  (\ref{LarDJ}). Thus the solution of the coupled
system  (\ref{LarDS}),  (\ref{LarDJ})   should as usual be sought in the form of the sum of the two terms depending on
time through the  factors $\sin \w_{\mathrm{L}} t$ and $\cos\w_{\mathrm{L}}t$.

According to what was said in the body of the paper, we seek for a solution of the system
\bea
\p_t{\D\f S} &+& \p_z\f J_z=0,\label{oscSz}\\
\p_t{\D\f J_z} &+& \F{w^2}3(\p_z\f S +\F{gE}{c}(\h{\f x}\sin \w_{\mathrm{L}} t-\h{\f y}\cos\w_{\mathrm{L}}t)\X(\f S-\f\w_{\mathrm{L}}))+ \kappa\D\f S\X\f
J_z=-\F{\D\f J_z}{\tau_1}\label{oscJz}
\eea
in the form $\D\f J_z=(D_0 S_0/L)(\f a \sin \w_{\mathrm{L}} t+\f b
\cos\w_{\mathrm{L}}t)$. From the first equation we get $\D\f S=(D_0 S_0/L)(-(\f b'/\w_{\mathrm{L}})\sin \w_{\mathrm{L}} t+(\f a'/\w_{\mathrm{L}})\cos\w_{\mathrm{L}}t)$, where
$\f a'=\p_z\f a$.

Next we should utilize the boundary condition of the absence of spin transfer through the walls of the channel: $\D\f J_z=0$
at the walls. Hence the solution can be expanded into harmonic modes, in which $\D\f J_z$  depend on $z$ through the
factors of the form $\sin kz$. Here $k=\pi{\mathsf Z}/d$ and $d$ is the thickness of the channel, which in general is  a function of
$x$.

Then the amplitudes of the modes in the simpler case of $|v|\le\pi/2$ will be
\be
\f a=\F{\e (s-\w_{\mathrm{L}}/S_0)}{C^2+1}(C\h{\f x}+ \h{\f y}),\qquad \f b=\F{\e (s-\w_{\mathrm{L}}/S_0)}{C^2+1}(\h{\f x}-C \h{\f y}),
\ee
where $C=\w_{\mathrm{L}}\tau_1-D_0k^2/\w_{\mathrm{L}}-vs$. We  here allowed for the fact that  due to diffusion after some finite time after
the beginning of the experiment  $S_0$ is not equal to $\w_{\mathrm{L}}$.

\end{document}